# Upper-division Student Understanding of Coulomb's Law: Difficulties with Continuous Charge Distributions

Bethany R. Wilcox\*, Marcos D. Caballero\*, Rachel E. Pepper† and Steven J. Pollock\*

\*Department of Physics, University of Colorado Boulder
†Departments of Integrative Biology and Civil & Environmental Engineering, University of California Berkeley

Abstract. Utilizing the integral expression of Coulomb's Law to determine the electric potential from a continuous charge distribution is a canonical exercise in Electricity and Magnetism (E&M). In this study, we use both think-aloud interviews and responses to traditional exam questions to investigate student difficulties with this topic at the upper-division level. Leveraging a theoretical framework for the use of mathematics in physics, we discuss how students activate, construct, execute and reflect on the integral form of Coulomb's Law when solving problems with continuous charge distributions. We present evidence that junior-level E&M students have difficulty mapping physical systems onto the mathematical expression for the Coulomb potential. Common challenges include difficulty expressing the difference vector in appropriate coordinates as well as determining expressions for the differential charge element and limits of integration for a specific charge distribution. We discuss possible implications of these findings for future research directions and instructional strategies.
Keywords: electricity and magnetism, Coulomb's Law, student difficulties with mathematics, electric potential
PACS: 01.40.Fk, 01.30.lb

## Introduction

Previous work has demonstrated that physics students struggle to manipulate and make sense of the mathematics necessary to solve problems in physics [1]. Particularly in upper-division courses, which require increasing mathematical sophistication, our community must move away from merely noting students' conceptual difficulties and focus on students' difficulties using mathematics in ways that are integrated with their conceptual understanding. Developing students' ability to connect mathematical expressions to physics concepts is often an explicit goal for these more advanced courses. For example, consensus learning goals [2] for upper-division courses at University of Colorado Boulder (CU) include "Students should be able to translate a physical description of an upper-division physic problem to a mathematical equation necessary to solve it," and "Students should be able to achieve physical insight through the mathematics of a problem."

Direct calculation of the electric potential or electric field from a continuous charge distribution by Coulomb's law is one promising topic for exploring student difficulties with mathematics in physics at the upper-division level. Students' conceptual difficulties with the electric field, potential, and Gauss' Law have been well documented [3,4]. Additionally, several Tutorials on this topic have been developed for both introductory and advanced courses [5,6]. However, investigations of specific student difficulties with the integral expression for Coulomb's Law are rare. This paper leverages a theoretical framework which seeks to identify which mathematical resources are Activated, Constructed, Executed, and Reflected upon (ACER) to frame the results of an investigation into student difficulties with Coulomb's Law. We use Coulomb's Law to refer to the integral equation allowing for direct calculation of the electric field or potential from a continuous charge distribution.

## The ACER Framework

A significant focus of this is study to investigate how and when students successfully connect the mathematical formalism of Coulomb's Law with the physics of a particular problem. At the introductory level several frameworks for the use of mathematics in physics have been developed by coordinating multiple theories of learning [7], adapting frameworks from mathematics education research [8], and providing a structure for physics problem solving [9].

Caballero et al. [10] leverage task analysis [11] and resource theory [12] to construct the ACER framework for investigating students' use of mathematics in upper-division physics courses where mathematical and conceptual difficulties are not always readily distinguishable. Implementation of the framework involves the creation of a researcher-guided outline of key elements in a correct solution. This framework is related to previous work from both Redish and Heller et al. [1,9]; however, the ACER framework is not a model for student reasoning or a rubric for student problem solving. Instead, it allows an instructor or researcher to pinpoint student difficulties by providing a scaffold for organizing the elements of a student's solutions. As a demonstration of the value of this framework, we discuss its application to problems determining the potential or electric field from a static, arbitrary charge distribution. We will focus only on problems that cannot be solved using Gauss' Law.

*Activation of the tool:* The first component of the framework involves the selection of a solution method. We identify four cues that are likely to activate resources identifying direct integration (i.e., Coulomb's Law) as the appropriate tool.
- The problem asks for the potential or electric field.
- The problem gives a charge distribution.
- The charge distribution does not have appropriate symmetry for the direct use of Gauss' Law.
- Direct calculation of the quantity is most efficient.

This fourth point is included to account for the possibility of solving for potential by first calculating the electric field or vice-a-versa. These methods are valid but often considerably more difficult.

*Construction of the model:* This component of the framework deals with constructing the math-physics connection. Here, both math and physics resources are used to map the specific physical situation onto the general mathematical expression for Coulomb's Law. The resulting integral expression should be in a form that could be solved with no knowledge of the physics of this specific problem. We identify four key elements that must be completed in this mapping.
- Use the geometry of the charge distribution to select a coordinate system.
- Express the differential charge element (dq) in the selected coordinates.
- Select limits consistent with the differential charge element and the extent of the physical system.
- Express the difference vector ($\vec{r} - \vec{r}\,'$, Griffiths' script-r) in the selected coordinates.

The second and fourth elements can be accomplished in multiple ways, often involving several smaller steps. In order to express the differential charge element, the student must combine the charge density and differential to produce an expression with the dimensions of charge (e.g., $dq = \sigma\, dA$). Construction of the difference vector often includes a diagram that identifies vectors to the source point and field point.

*Execution of the mathematics:* This component of the framework deals with the mathematics required to compute a final expression. In addition to using a variety of mathematical resources (e.g., manipulating algebraic expressions and executing multivariable integrals), this computation requires an awareness of which variables are being integrated over, as there are two sets (one for each the source point and field point).

*Reflection on the result:* The final component of the framework involves verifying that the expression is consistent with expectations. While many different techniques can be used to reflect on the result, these two checks are particularly common:
- Verify that the units are correct.
- Check the limiting behavior to ensure it is consistent with the total charge and geometry of the charge distribution.

## Methods

Data for this study were collected during the first semester of the junior-level Electricity and Magnetism (E&M) sequence at CU. This course consistently covers the first 6 chapters of Griffiths' textbook [13] which includes both electrostatics and magnetostatics. The student population is composed of physics, engineering physics, and astrophysics majors in their junior or senior year, and a typical class size is 30-60 students. The E&M 1 course at CU has been transformed to include a number of research-based techniques including peer instruction [14] using clickers and optional weekly tutorials [6]. In order to determine the types of difficulties students have with Coulomb's Law, we analyzed student solutions to a canonical exam problem on continuous charge distributions and conducted five, formal think-aloud interviews to further probe student understanding.

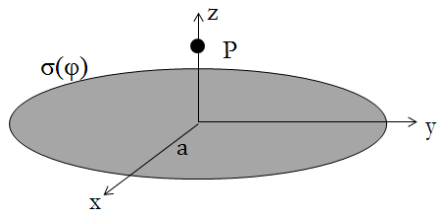

FIGURE 1. An example of the canonical exam problem on continuous charge distributions. In this case, the student would be prompted to calculate the electric potential at point P on the z-axis from a disk with surface charge density $\sigma(\varphi)$.

In the exam problem, students were asked to calculate the electric potential along an axis of symmetry from a disk with charge density $\sigma(\varphi)$ (see Figure 1). Our students were exposed to the Coulomb's Law integral for the electric field (E) before the analogous expression for the electric potential (V). However, the vector nature of the electric field makes E significantly more challenging to calculate and, historically, instructors at CU ask for the potential on exams. Exams were collected from four semesters of the course (N=172), each taught by a different instructor. Two of these instructors were physics education researchers involved in developing the transformed materials and two were traditional research faculty. All four semesters utilized some or all of the available transformed materials. The exact details of the disk question, while very similar, were not identical from semester to semester. Interview participants were paid volunteers who had successfully completed E&M 1 one semester prior. The students were asked to solve a slightly more advanced version of the exam problem where they attempted to calculate the potential from two disks while talking through their reasoning. The integral formulae for the potential and electric field from an arbitrary charge distribution were provided in notation consistent with that from Griffiths [13].

Exams were analyzed by identifying each of the key elements from the framework that appeared in the students' solution. Each element was then coded to identify the types of steps made by students. The interviews were similarly analyzed by classifying each of the student's major moves into one of the four components of the framework. In this way, exams provided quantitative data identifying common difficulties and interviews offered deeper insight into the nature of those difficulties. The initial interview protocol was developed to probe the preliminary difficulties identified in the student exam solutions. At this point, the ACER framework had not been developed, and as a result, the interview protocol provided no data relevant to Activation.

**Results and Discussion**

Preliminary analysis of the exams and interviews was difficult as many students make multiple mistakes which then carry throughout the solution. The ACER framework provided an organizing structure by focusing on important nodes in students' solution instead of attempting to classify the entire solution.

*Activation of the tool:* Identifying evidence of Activation in the exam solutions was challenging as the students did not typically write out their thought process as they began the problem. However, roughly a quarter of the students (27%, N=46) attempted to calculate V by first calculating E and then taking a line integral. These students did not recognize that a direct calculation of V is more efficient. Roughly half of this group (54%, N=25) used the Coulomb's Law integral for the electric field to calculate E, and the other half (46%, N=21) attempted to employ Gauss' Law. Students in the latter group justified their answers with comments such as, "Since we want the voltage at a point outside the disk, the E-field we use will appear to be that of a point charge at the origin," suggesting that they did not recognize that a finite disk symmetry is not appropriate for Gauss's Law.

In three of the four semesters the students were asked to calculate the total charge on the disk before they were asked for the potential. In the remaining semester (N=25), the students were only asked to describe the angular charge distribution. No students in this semester attempted to use Gauss' Law to solve for the potential. This may indicate that calculation of the total charge first activates resources that encourage students to use Gauss' Law.

*Construction of the model:* Of the 172 exams, 154 contained elements coded as belonging to the Construction component. Only two of these students did not use cylindrical variables, indicating that the students at this level are adept at selecting an appropriate coordinate system in highly symmetric problems. However, when setting up the integral, nearly half of the students (41%, N=64) had difficulty expressing the differential charge element, some (13%, N=20) provided limits of integration that were inconsistent with the differential, and about half (47%, N=72) incorrectly expressed the difference vector. The most common errors made while expressing the differential charge element (dq) were (see Table 1): plugging in the total charge instead of the charge density, performing the integration over a region of space where the charge density was zero, and not including the cylindrical Jacobian. There was greater variability in the expressions for the difference vector (script-r), but some of the common errors included (see Table 2): using the expression for the magnitude appropriate for a ring of charge, setting the magnitude equal to the distance to the source point, and setting the magnitude equal to the distance to the field point.

Initially, we interpreted these difficulties as a failure to conceptualize the integral as a sum of the contributions to V from each little 'bit' of charge, dq. However, the interviews suggest that the problem may be more subtle than that. Instead, these difficulties center on students' ability to synthesize conceptual and mathematical resources to construct mathematical expressions consistent with the physical descriptors of the problem. All five interview participants drew vectors that could represent the difference vector on their diagrams, and four of them explicitly identified these vectors as script-r. Yet only one of these students was able to produce the correct expression for script-r, while the others made errors similar to those listed in Table 2. Additionally, four out of five of the interview participants made statements and/or gestures indicating they understood the integral to be a sum over the charged disk. Furthermore, post-test data from the first semester of the Classical Mechanics and Mathematical Methods course at CU shows that more than 80% of students can correctly determine the differential area element for a cylindrical shell, when asked directly, one semester prior to taking E&M. Yet three of these four interviewees made mistakes similar to those listed in Table 1.

**TABLE 1.** Difficulties expressing *dq*. Percentages are of just the students who had difficulty with *dq* (41%, N=64). Codes are not exhaustive or exclusive.

| Difficulty | N | Percent |
|---|---|---|
| Calculated total charge<br>e.g. $dq = Q_{tot}\, dr\, dz\, r d\varphi$ | 10 | 16 |
| Not integrating over charges<br>e.g. $dq = \sigma\, dr\, dz\, r d\varphi$ | 39 | 61 |
| Differential w/ wrong units<br>e.g. $dq = \sigma\, dr\, d\varphi$ | 12 | 19 |
| Other | 10 | 17 |

**TABLE 2.** Difficulties expressing *script-r*. Percentages are of just the students who had difficulty with *script-r* (47%, N=72). Codes are not exhaustive.

| Difficulty | N | Percent |
|---|---|---|
| Ring of charge<br>e.g. $script\text{-}r = \sqrt{a^2 + z^2}$ | 27 | 38 |
| Distance to the source point<br>i.e. $script\text{-}r = r'$ | 11 | 15 |
| Distance to the field point<br>i.e. $script\text{-}r = z$ | 10 | 14 |
| Never expressed *script-r* | 9 | 12 |
| Other | 15 | 21 |

*Execution of the mathematics:* Overall, 103 exam solutions contained elements from the Execution component, because one of the four instructors (N=55) only asked the students to set up the integral. Of these students, half (51%, N=53) made mathematical errors while solving integrals in the disk problem.

Fewer than half of the students with mathematical errors (41%, N=22) made only slight math errors (i.e., dropping a factor of two or plugging in limits incorrectly). The rest (59%, N=31) made significant mathematical errors (i.e., pulling integration variables outside of integrals and/or not completing one or more integrals). It is worth noting that roughly a quarter of the students made significant math errors, but the significant variation in the types of mistakes and the timed nature of the exams limits our insight into why a particular error was made. Though the interviews were not structured to emphasize the execution component, four of five students attempted to compute integrals. Two made only slight math errors and the remaining two made significant mathematical errors.

*Reflection on the Result:* In most cases, mistakes in the Construction or Execution component resulted in expressions for the potential which had the wrong units and/or limiting behavior. However, we found that without being prompted, our students did not check these properties to gain confidence in their solutions. Only a small number of students (8%, N=13) made explicit attempts to check their solutions on exams and only by checking the limiting behavior. Three interview participants made no attempt to reflect on their solutions, while the other two made superficial statements about whether the solution looked familiar.

**Summary and implications**

This paper presents an application of the ACER theoretical framework to guide the analysis of data on student difficulties with direct calculation of the potential from a continuous charge distribution in upper-division E&M. We find that our junior-level students have difficulty, (1) activating Coulomb's Law as a solution method, (2) coordinating mathematics and physics resources, and (3) spontaneously reflecting on the result. We consider the second and third difficulties to be most important because the ability to translate between physical and mathematical descriptions of a problem and to reflect meaningfully on the result are two defining characteristics of a physicist. To address these difficulties, we advocate an increased focus on teaching students to construct the differential charge element and difference vector, as well as explicit discussion of how and when to reflect on their solutions. We also encourage instructors to be aware of the resources their prompts cue students to access.

The ACER framework revealed the need for additional interviews with a new protocol that targets both Activation and Reflection, probes more deeply into how students synthesize their conceptual and mathematical resources, and investigates students' mathematical difficulties with multivariable integrals. The ACER framework provided an organizational structure which pinpointed student difficulties with a canonical E&M problem involving sophisticated mathematics and physics. The framework is being refined and future work will include implementing ACER to structure investigations of upper-division student difficulties from the ground up.

This work is funded by NSF grant DUE-1023028.

**References**
1. E. Redish, In Conf. Proc. World View on Physics Education in 2005: Focusing on World change (2009)
2. R. Pepper et al., 2011 PERC proc. 1413, 291-294
3. R. Pepper et al., Phys. Rev. ST PER 8, 010111 (2012)
4. C. Singh, Am. J. Phys. 74(10), (2006)
5. Z. Isvan et al., 2011 PERC proc. 883, 181-184
6. S. Chasteen et al., 2008 PERC proc. 1064, 91-94
7. J. Tuminaro, Dissertation, UMD, College Park
8. C. Rebello et al., 2011 PERC proc. 1413, 311-314
9. K. Heller et al., The competent problem solver, McGraw-Hill (1995)
10. M. Caballero et al., submitted to 2012 PERC proc.
11. R. Catrambone, J. Exp. Psychol. Learn, 22, 1020 (1996)
12. D. Hammer et al., Transfer of learning from a modern multidisciplinary perspective, pp. 89-120 (2005)
13. D. J. Griffiths, Introduction to Electrodynamics 3rd Ed., Prentice Hall (1999)
14. E. Mazur, Peer Instruction, Prentice Hall (1997)